\documentclass[%
aps,reprint,
 amsmath,amssymb,
]{revtex4-2}  

\usepackage{graphicx}
\usepackage{dcolumn}% Align table columns on decimal point
\usepackage{bm}% bold math
\usepackage{todonotes}
\usepackage{hyperref} 

\usepackage{caption}
\usepackage{overpic} 
\usepackage{color,soul}
\usepackage{float}

\usepackage{hyperref} 
   
\usepackage{subfig} 

\definecolor{orange2}{RGB}{190, 78, 50}

\definecolor{grau}{RGB}{100 100 100}
\definecolor{grau2}{RGB}{220 220 220}
\definecolor{blau}{RGB}{68 107 131}

\DeclareFontFamily{U}{mathx}{}
\DeclareFontShape{U}{mathx}{m}{n}{<-> mathx10}{}
\DeclareSymbolFont{mathx}{U}{mathx}{m}{n}
\DeclareMathAccent{\widehat}{0}{mathx}{"70}
\DeclareMathAccent{\widecheck}{0}{mathx}{"71}

\begin{document} 

\title{High-fidelity experimental model verification for flow in fractured porous media}
\thanks{This work is funded by the UoB Akademia-project \textit{FracFlow}.}

\author{Jakub W.\ Both, Eirik Keilegavlen and Jan M. Nordbotten}
\email{jakub.both@uib.no}
\affiliation{Centre for Modeling of Coupled Subsurface Dynamics, Department of Mathematics, University of Bergen, Norway}

\author{Bergit Brattekås and Martin Fernø}
\affiliation{Department of Physics and Technology, University of Bergen, Norway}

\date{\today}

\begin{abstract}
Mixed-dimensional mathematical models for flow in fractured media have been prevalent in the modeling community for almost two decades, utilizing the explicit representation of fractures by lower-dimensional manifolds embedded in the surrounding porous media. 
In this work, for the first time, direct qualitative and quantitative comparisons of mixed-dimensional models are drawn against laboratory experiments. Dedicated displacement experiments of steady-state laminar flow in fractured media are investigated using both high-resolution PET images as well as state-of-the-art numerical simulations.
\end{abstract}

\maketitle

The presence of fractures strongly influences both flow and transport in porous media. As fractures are ubiquitous in many geological rocks \cite{berkowitz2002characterizing} and are induced by subsurface operations \cite{olasolo2016enhanced}, accurate, reliable, and verified models for fractured porous media are essential in the modeling and simulation of flow and transport in fractured porous media subsurface.  

Since their inception about 20 years ago~\cite{alboin2002modeling,martin2005modeling,angot2009asymptotic}, models where fractures are represented as lower-dimensional objects (relative to the surrounding rock), have received much attention. 
Such \textit{mixed-dimensional fracture models}, as we will refer to them, provide a natural framework for modeling and efficient computation related to fractured porous media. The models have therefore been extensively studied both in terms of their approximation properties to full \textit{equi-dimensional models}, see, e.g.,~\cite{martin2005modeling} as well as their mathematical properties, see, e.g.,~\cite{boon2021functional}. In numerical benchmark studies, mixed-dimensional models, as well as their numerical discretization, have been further validated against numerical discretizations of equi-dimensional models~\cite{flemisch2018benchmarks}. 

Flow experiments have been key to understanding the process of fluid transfer between fracture and matrix, and between adjacent matrix blocks: often focusing on multi-phase flow and the balance of capillary to viscous and gravitational forces~\cite{rangel2002experimental}. Important recovery mechanisms in fractured media, such as spontaneous imbibition~\cite{MASON2013268} and gravity drainage~\cite{Firoozabadi2000} have been investigated experimentally. \textit{In situ} imaging has occasionally been applied to improve insight, e.g. revealing the existence of wetting-phase bridges forming in vertical fractures to aid capillary continuity~\cite{aspenes2008wetting}.

Despite the rich literature both from the modeling and experimental perspectives, direct validation of mixed-dimensional fracture models as compared to actual physical flow in fractured rock is largely missing. As a consequence, key questions regarding the model applicability have not been addressed, in particular as relates to fracture tips and intersections. Indeed, the majority of modeling literature has emphasized the role of the fracture and its interaction with the matrix, and the correct modeling of fracture tips and intersections is often either treated summarily~\cite{boon2018robust,berre2019flow}, or simply assumed to not exist in the sense that fractures are assumed to extend to the boundary of the domain~\cite{martin2005modeling}. In the context of discrete fracture networks, which follow a different methodology, intersections have been discussed, partially with dependence on the intersection angle~\cite{schwenck2015dimensionally}. Considering this background, we identify three key objectives, which to our knowledge have not been satisfactorily addressed in previous works:  

Primarily, we ask: Is mixed-dimensional modeling of fractured porous media a suitable framework for quantitative analysis?

Our primary objective is substantiated through two secondary objectives. 
Most concretely, we address: Does the actual physical geometry of the fracture tip impact the flow both in the fracture and surrounding matrix? From a modeling perspective, mass balance arguments imply a no-flow condition on the fracture tip~\cite{angot2009asymptotic}. 
Other works have discussed non-trivial pressure and flux singularities at fracture tips depending on the fracture tip geometry~\cite{chen2013pressure} as well as imposing a matching flux coupling through an additional degree of freedom~\cite{schwenck2015xfem}, postulated to be of larger relevance in the case of high transversal permeability. The above references notwithstanding, the most common modeling choice is a flow barrier at the fracture tips. As the real physical system has no flow barrier at fracture tips, this implies a modeling assumption that the tip geometry is completely irrelevant. 

Implicitly, we also address: Is the modeling of fracture intersections of suitable accuracy? In the modeling literature, various treatments of  fracture intersections have been introduced, where in a conceptual sense the extreme cases are pressure continuity~\cite{schwenck2015dimensionally}, and local flux laws~\cite{angot2009asymptotic}. To our knowledge, no rigorous justification for either position exists from a physical perspective, nor has it been established whether these nuances matter within the context of the overall uncertainties of subsurface flows.

\begin{figure}[!t]  
\includegraphics[width=0.45\textwidth]{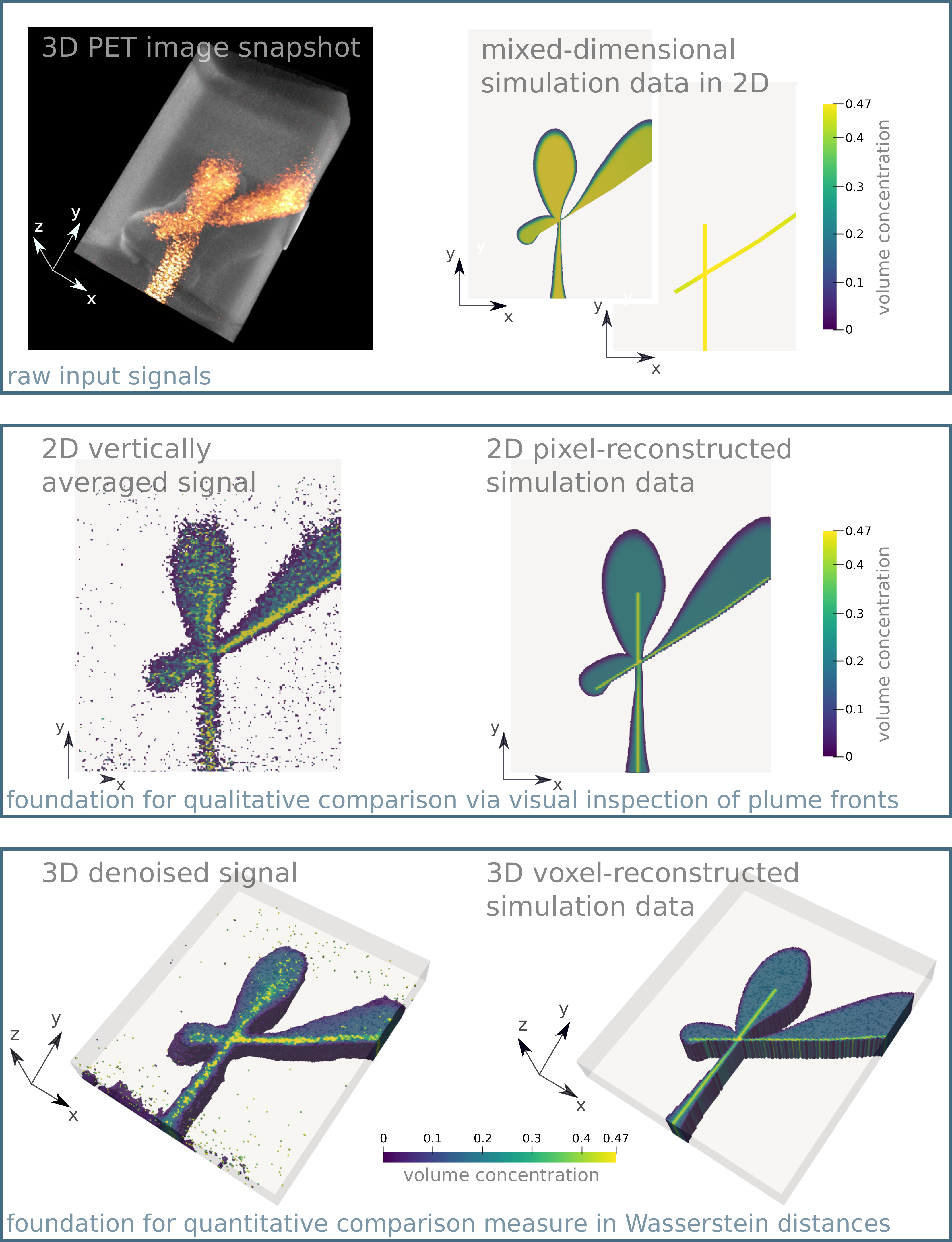} 
\caption{\label{fig:postprocessing-experimental} Schematic of available data formats considered in the validation study. PET images in vertically averaged 2D and denoised 3D views; mixed-dimensional simulation data in compatible equi-dimensional formats.}
\end{figure}

To address the key modeling questions identified above, we consider a series of qualitative and quantitative comparison studies between dedicated high-fidelity laboratory tracer experiments, visualized by high-resolution \textit{in situ} positron emission tomography (PET) imaging~\cite{Ferno2015}, and numerical mixed-dimensional fracture models utilizing the state-of-the art open-source fracture flow simulator PorePy~\cite{keilegavlen2021porepy} (both detailed below), illustrated in Fig.~\ref{fig:postprocessing-experimental}. 
The experiments are designed to enable the following discussion of the spatio-temporal data:
\begin{itemize}
\item A pair of experiments is constructed, only differing in the absence and presence of added physical flow barriers in the fracture tips, the latter resembling the mathematical modeling assumption, thus directly adressing the impact of the fracture tip.
 \item Disparities of tracer plumes between experiments and mixed-dimensional fracture models are compared qualitatively through visual inspection and quantitatively in terms of Wasserstein distances, considering experiments with and without fracture intersections, thus implicitly adressing the impact of the modeling of intersections.
\end{itemize}  

Before highlighting the different experiments of this study, let us recall the principles of mixed-dimensional modeling, which will be challenged by the presented study. The dimensional reduction and representation of fractures as lower-dimensional objects has three central implications: the introduction of (I) a mixed-dimensional geometry, (II) a mixed-dimensional representation of physical fields, and (III) meaningful model equations coupling the physics between adjacent objects of varying dimensions~\cite{martin2005modeling,boon2018robust,boon2021functional}. In the following, this is exemplified in the context of flow and tracer transport.

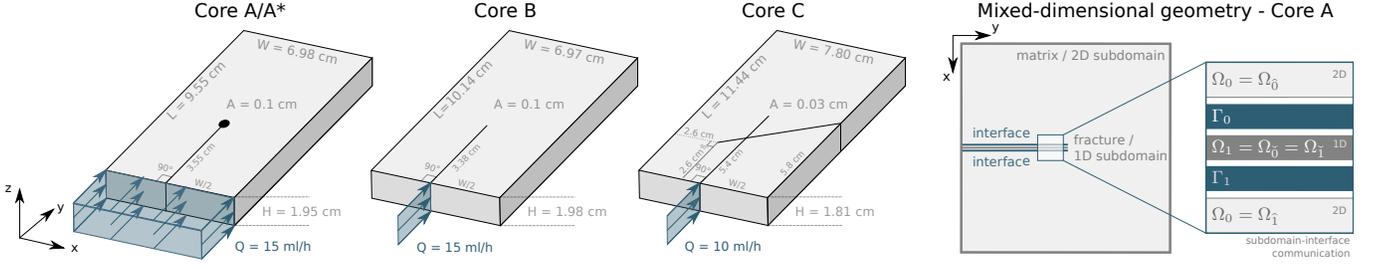
\begin{figure*}[!t]  
    \begin{overpic}[width=\textwidth]{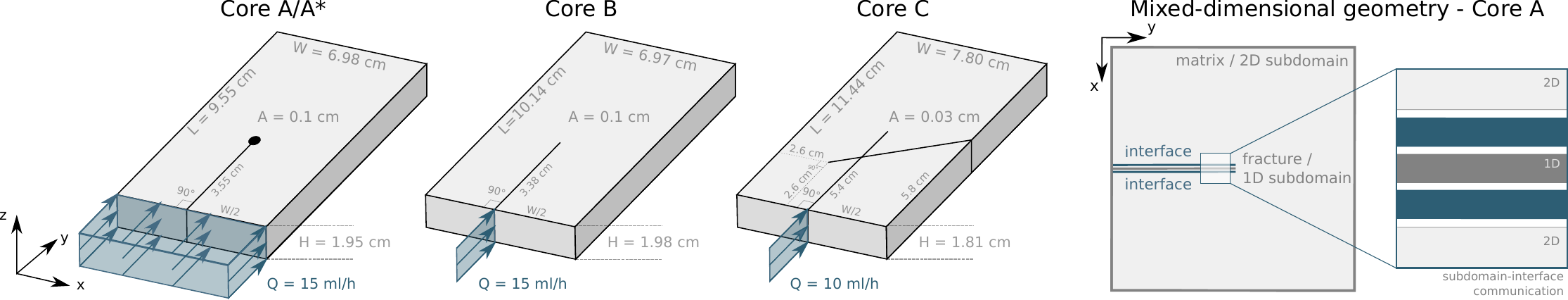}
    \put(89.5,13){\scalebox{0.75}{\textcolor{grau}{$\Omega_0 = \Omega_{\widehat{0}}$}}}
    \put(89.5,10.3){\scalebox{0.75}{\textcolor{white}{$\Gamma_0$}}}
    \put(89.5,8.0){\scalebox{0.75}{\textcolor{white}{$\Omega_{1} = \Omega_{\widecheck{0}} = \Omega_{\widecheck{1}} $}}}
    \put(89.5,5.7){\scalebox{0.75}{\textcolor{grau2}{$\Gamma_1$}}}
    \put(89.5,3){\scalebox{0.75}{\textcolor{grau}{$\Omega_0 = \Omega_{\widehat{1}}$}}}
\end{overpic}
\caption{\label{fig:experiments-geometry} \textit{Left:} Geometrical specifications of Cores A/A$^\star$ (A$^\star$ but with splinter in tip), B, and C, including length ($L$), width ($W$), height ($H$), fracture aperture ($A$); injection strategy and rate ($Q$) \textcolor{blau}{(blue)}. \textit{Right:} Conceptual mixed-dimensional representation of Core A in 2D with subdomains $\{\Omega_i\}_{i\in I}$ \textcolor{grau}{(gray)} and interfaces $\{\Gamma_j\}_{j \in J}$ \textcolor{blau}{(blue)}.}
\end{figure*}

(I) \textit{Mixed-dimensional geometry.} Fractures are three-dimensional geometrical features with high aspect ratios. Compared to the dimensions of the surrounding matrix, the width of their cross section is several orders of magnitude smaller. Hence, from a geometrical standpoint, a lower-dimensional representation of fractures appears natural. The argumentation continuous recursively to fracture intersections. Consequently, a general fractured media $\Omega\subset \mathbb{R}^N$ can be conveniently described as a hierarchy  of mixed-dimensional geometries, $\{\Omega_i\}_{i\in I}$, $I:= \{1,...,m\}$, with the matrix as subdomain of the ambient dimension, fractures of codimension 1 (planes), intersections of fractures of codimension 2 (lines), and again their intersections of codimension 3 (points)~\cite{boon2018robust}. Interfaces, $\{\Gamma_j\}_{j\in J}$, $J:=\{1,...,M\}$, between subdomains of co-dimension one allow for data exchange via suitable projections. Examples are displayed in Fig.~\ref{fig:experiments-geometry}.  

(II) \textit{Mixed-dimensional fields.} As a result of the dimensional reduction~\cite{martin2005modeling}, physical fields, originally defined in the equi-dimensional geometry, receive distinct representations on each subdomain, collected in the form of a mixed-dimensional field. In the context of flow and transport, the mixed-dimensional fluid pressure $\{p_i\}_{i\in I}$ and tracer concentration $\{c_i\}_{i \in I}$ are defined.

(III) \textit{Mixed-dimensional equations.} A complete mathematical description of flow on mixed-dimensional geometries requires the assignment of equations on both the subdomains and the connecting interfaces. Starting from an equi-dimensional model, these can be derived using averaging or integration over the fracture apertures. In addition, postulation of closure relations is required. A key feature of the prototypical model recalled below is the conceptually identical treatment of each dimension.

The focus of this study lies on incompressible, quasi-static, laminar flow in porous media. Based on the fundamental principle of mass conservation together with Darcy's law, the governing equation for flow on subdomain $\Omega_i$, $ i \in I$, is given by
\begin{align}
\label{eq:flow-subdomain}
    \nabla_{||}\cdot \left(-v_i \mu^{-1} K_{i,||}\nabla_{||}p_i \right) = f_i + {\textstyle \sum}_{j\in \widehat{S}_i} \lambda_j 
\end{align}
where $v_i$ is the specific volume resulting from dimension reduction (equals $1$ at ambient dimension, the fracture width for fractures, their product for intersections etc.), $\nabla_{||}$ is the tangential differential operator (void on 0-dimensional subdomains), $K_{i,||}$ is the tangential permeability tensor, $\mu$ is the dynamic viscosity, $f_i$ is the volumetric source, $\widehat{S}_i$ identifies neighbouring interfaces towards subdomains of one dimension higher (empty for subdomains of ambient dimension) and $\{\lambda_j\}_{j\in \widehat{S}_i}$ are associated interface fluxes. The tangential permeability $K_{||,i}$ simply equals the matrix permeability at ambient dimension, on fractures and intersections we choose a cubic law associated to perfect Poiseuille flow~\cite{berre2019flow}, i.e., we set $K_{||,i} = \tfrac{v_i^2}{12}$ for fractures.

On neighboring pairs of lower and higher dimensional subdomains, $\Omega_{\widecheck{j}}(=\Omega_i)$ and $\Omega_{\widehat{j}}$ (with $\widecheck{(\cdot)}$ and $\widehat{(\cdot)}$ associating subdomains with interface $(\cdot)$, cf.\ Fig.~\ref{fig:experiments-geometry}), the flux across their interface $\Gamma_j$ respectively appears as source terms as in~\eqref{eq:flow-subdomain} and Neumann boundary condition
\begin{align}
    \left(- \mu^{-1}\, K_{||,\widehat{j}} \nabla_{||} p_{\widehat{j}}\right) \cdot n_{\widehat{j}} = \lambda_j\quad\text{on }\partial\Omega_{\widehat{j}}\cap \Gamma_j
\end{align}
where
$n_{\widehat{j}}$ is the outer normal vector onto $\partial\Omega_{\widehat{j}}\cap \Gamma_j$. Based on a linearity assumption, the flux $\lambda_j$ across $\Gamma_j$ follows a Darcy-like law, introducing the normal permeability $\kappa_{j}$, inversely correlated to the intrinsic aperture $a_j$ ~\cite{boon2018robust}), effectively representing a discrete normal derivative~\cite{martin2005modeling},
\begin{align}
 \label{eq:normal-permeability}
    \lambda_j &= -\mu^{-1} \, \kappa_{j} \left(p_{\widehat{j}} - p_{\widecheck{j}} \right),\qquad   
    \kappa_j = v_{\widehat{j}} K_{||,\widecheck{j}} / (\tfrac{1}{2} a_j). 
\end{align} 
Boundary conditions are assigned to close the system. At the external boundaries of the host medium $\Omega$, pressure and flux boundary conditions depend on the use case. At internal boundaries, i.e., fracture tips, we employ the widely used consensus and impose \textit{no-flow conditions}
\begin{align}
\label{eq:no-flow-bc}
    \left(-\mu^{-1} \, K_{||,i} \nabla_{||} p_i \right) \cdot n_i = 0.
\end{align}
With these modeling choices, all parameters are defined exclusively in terms of the matrix permeability and fracture apertures, and thus our model as applied in this study contains no free tuning parameters. 

Following the same methodology, passive tracer transport in fractured media can be similarly modelled as mixed-dimensional advection-dispersion~\cite{fumagalli2013reduced}. The tracer transport in subdomain $\Omega_i$, $i\in I$, is governed by
\begin{align}
\label{eq:transport-subdomain}
v_i\left[\phi_i \partial_t c_i + \nabla_{||} \cdot \left( c_i q_i - D_{||,i} \nabla_{||} c_i \right) \right] = h_i + \sum_{j\in \widehat{S}_i} \eta_j,
\end{align}
where $\phi_i$ is the porosity,
$q_i = -\mu^{-1} \,K_{||,i}\nabla_{||}p_i$ is the advective flux,  $D_{||,i}$ is the tangential dispersion (tensor), and $h_i$ is the volumetric source term. In this work, the dispersion $D_{||,i}$ follows an anisotropic model, decomposing in longitudinal dispersion $D_{\mathrm{L},||,i} = \alpha_\mathrm{L} |q_i|$ and transversal dispersion $D_{\mathrm{T}, ||,i} = \alpha_\mathrm{T} |q_i|$, see e.g.~\cite{de1986quantitative}. 
The system is closed following the same modelling principles as for mixed-dimensional flow, introducing advective-diffusive interface fluxes $\eta_j$. 

Numerical simulations allow for approximating mixed-dimensional models, such as~\eqref{eq:flow-subdomain}--\eqref{eq:transport-subdomain}, in particular for highly complex geometries. For this work, the choice of the numerical discretization is not essential, as long as it is locally conservative, consistent and stable, across all dimensions. This ensures that by choosing a sufficiently fine grid, the approximation error is negligible as compared to the modeling error. 
In this study, a locally conservative finite volume discretization is employed with diffusive fluxes approximated with the MPFA method~\cite{aavatsmark2002introduction, nordbotten2021introduction}, yet tailored to mixed-dimensional models following the unified framework in~\cite{nordbotten2019unified}. Convective fluxes are approximated using first-order upstream weighting~\cite{eymard2000finite}. For the implementation the open-source software framework PorePy~\cite{keilegavlen2021porepy} is used, which provides mass-conservative finite volume discretizations that have been extensively validated against the model equations through participation in code comparison studies~\cite{berre2021verification}, and by a posteriori error analysis~\cite{varela2022posteriori}.

For the subsequent model validation study aiming at addressing the above research questions, dedicated laboratory tracer experiments are conducted.
For all four setups, illustrated in Fig.~\ref{fig:experiments-geometry}, cuboid Bentheimer sandstone core material was used and cut using a diamond saw blade to create thin, smooth fractures. Aiming at investigating fracture tips and intersections, the four configurations are chosen: two geometries with a simple cut (Core A and Core B); Core A but with the fracture tip closed by a metal splinter (Core A$^\star$) resembling an actual physical flow barrier and thus imitating the mathematical model equation~\eqref{eq:no-flow-bc}; and a geometry with a simple fracture network consisting of two intersecting fractures (Core C), realized by assembling the rock pieces with epoxy on a fixed plate. 
Each of the considered core samples is treated homogeneous and isotropic. Based on independent displacement experiments using intact core material, effective hydraulic matrix properties (porosity $\sim$ 0.2, permeability $\sim$ 2 D, dispersion $\sim$ 7e-4 m) are determined, cf.~\cite{Both2023supp} for detailed values.

The rock cores were initially saturated with 3.5\% NaCl brine (viscosity $\mu=$1.09cP at ambient temperature) under vacuum. Brine was thereafter injected into the inlet of each core, using slightly different conditions, cf.\ Fig.~\ref{fig:experiments-geometry}, considering injection across the entire inlet or directly into the fracture, while the outlet end face is open to flow and fluid is produced at atmospheric pressure. Polyoxymethylene end pieces were machined and attached to inlet and outlet end faces to define and facilitate injection and extraction. The remaining core faces are impervious to fluid flow through use of epoxy.

High-resolution \textit{in situ} PET imaging is used to track flow patterns, cf.\ Fig.~\ref{fig:postprocessing-experimental}. For this, the cores are placed in a high-resolution, multimodal PET-CT scanner and radioactive $^{18}$F-FDG-labelled brine is used as injection fluid. Low concentrations, still detectable by PET, are used to not notably alter fluid density and viscosity. As consequence the experiments are quasi-2D.  %The merely weak gravity effects result in horizontally symmetric fluid flow.

The foundation of the model validation study is the direct comparison of laboratory PET data and corresponding mixed-dimensional simulation data. Yet, such comparison solicits a common ground, here chosen as time-series of volumetric (Darcy-scale) concentration with respect to the matrix, cf.\ Fig.~\ref{fig:postprocessing-experimental}. %The advantage of such concentration data is that it allows for both qualitative and quantitative analysis. 

While simulation data undergoes a mass-conservative equi-dimensional reconstruction, the sparse PET signal of the experimental data requires space-time smoothing to extract Darcy-scale concentration data. To this end, snapshots are obtained from the dynamic 4D PET images through signal accumulation over 4D space-time cubes (60 seconds $\times$ 0.4 mm voxels); total variation denoising is used for simultaneous shape-preserving denoising and inpainting, and signal rescaling allows for matching the known injection rate, finally defining the cleaned data sets used in this study. This workflow retains the quality of the data set and regularizes the signal merely on the order of the measurement error~\cite{Both2023supp,nordbotten2023darsia}. Dimension reduction (3D to 2D) is applied to enable visual comparison of the (almost) plane symmetric tracer plumes. All data processing and analysis uses the open-source Darcy-Scale Image Analysis toolbox DarSIA~\cite{nordbotten2023darsia}.

Additionally to assessing the qualitative match via visual comparison of tracer plumes, the disparity between concentration data can be quantified in terms of the \textit{1-Wasserstein distance}~\cite{Both_wip} (Earth Mover's Distance~\cite{rubner1998metric}). It explicitly measures the mass-distance required to transport a tracer density from one location to another in form of a mass-conservative flux field (here termed \textit{Wasserstein flux}). For universal interpretation, we employ a relative distance with the reference value given by the average mass-distance of the current concentration profile from the injection boundary.

To obtain a baseline estimate of the deviation of the experiment from its design, we consider the deviation in the tracer distributions for Cores A, A$^\star$, and B, from a symmetric distribution (the experimental design has a North-South symmetry axis), in terms of a (relative) Wasserstein distance. The identical operating conditions across all experiments justify its interpretation as proxy for measurement uncertainty and the reduction to a characteristic value (of the order of 0.046 $\pm$ 0.013~\cite{Both2023supp}). We emphasize that this should be seen as a lower bound on the actual experimental uncertainty, as it does not account for any experimental artefacts that are not symmetry-breaking (impact of epoxy casing, imperfections in cutting of fracture, variability in the flow pumps, etc.).   

The access to experimental data~\cite{data} and corresponding accurate numerical simulation data~\cite{code} allows for model validation, divided as follows into two  sections.

\begin{figure}[H]
    \centering  
    \includegraphics[width=0.4\textwidth]{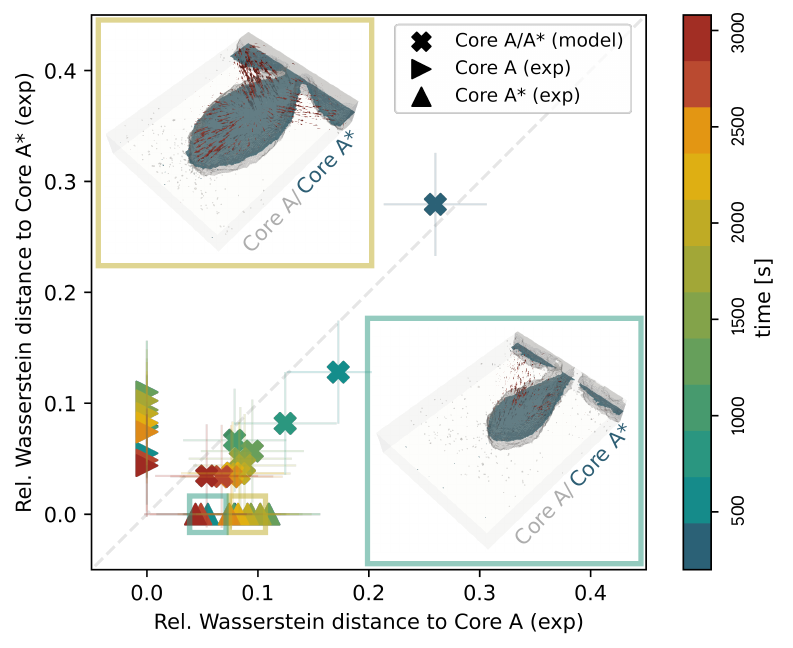}
    \caption{Quantitative cross comparison over time (color labeled) between experimental and simulation data for Cores A/A$^\star$; error bars indicate mean measuring uncertainty; Wasserstein fluxes between physical experiments.} 
    \label{figure:block-b-comparison}
\end{figure}

\begin{figure*}[!t]  
 \includegraphics[width=\textwidth]{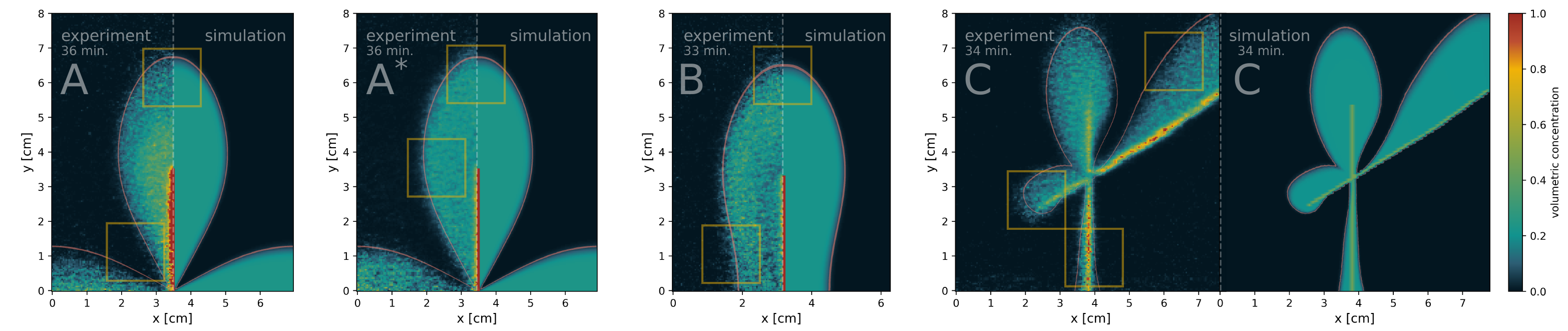}
 \caption{Concentration plumes for the experimental and simulation data in side-by-side comparison for Cores A, A$^\star$, B, C with yellow boxes putting emphasis on regions with minute discrepancies, further discussed in the text; additional red 5\% contour line of the simulated tracer to aid visual comparison; further snapshots displayed in~\cite{Both2023supp}. \label{fig:qualitative-comparison-late-time}}
\end{figure*}

\textit{Direct assessment of the impact fracture tip geometry} To probe the validity of the central modeling assumption of a no-flow condition in the fracture tip~\eqref{eq:no-flow-bc}, the main focus lies on comparing the laboratory experiments conducted using the Cores A and A$^\star$, recalling that these are identical from a modeling perspective. The Wasserstein distance for different time steps between the experimental data of the two different experiment series, as well as to the simulation data is employed to assess the fit. The relative distances are displayed in Fig~\ref{figure:block-b-comparison}, together with the Wasserstein flux for a single time step illustrating the structure of the error. Considering first only the experimental results (triangles), we observe that after an early-time transient, the deviation between the two experiments is on the order of 3-10\%, which is comparable to the experimental uncertainty. Contrasting the experiments with the simulation results, (indicated by crosses), we see that (to within experimental uncertainty) the mathematical model is equally close to either of the two experiments, and at later times is as close to the experiments as the variability between them. Seen together, we draw two observations: 

\begin{enumerate}
    \item The equidistance between the simulation and the two experiments indicates that the dominating modeling error cannot be ascribed to the treatment of the fracture tip. 
    \item Clustering of the experiments and the simulation within 5-10\% relative error rather implies that the modeling error is within the general uncertainty associated with the reproducibility of the experiment. 
\end{enumerate}
 
\begin{figure}[!h]
    \centering  
    \includegraphics[width=0.40\textwidth]
    {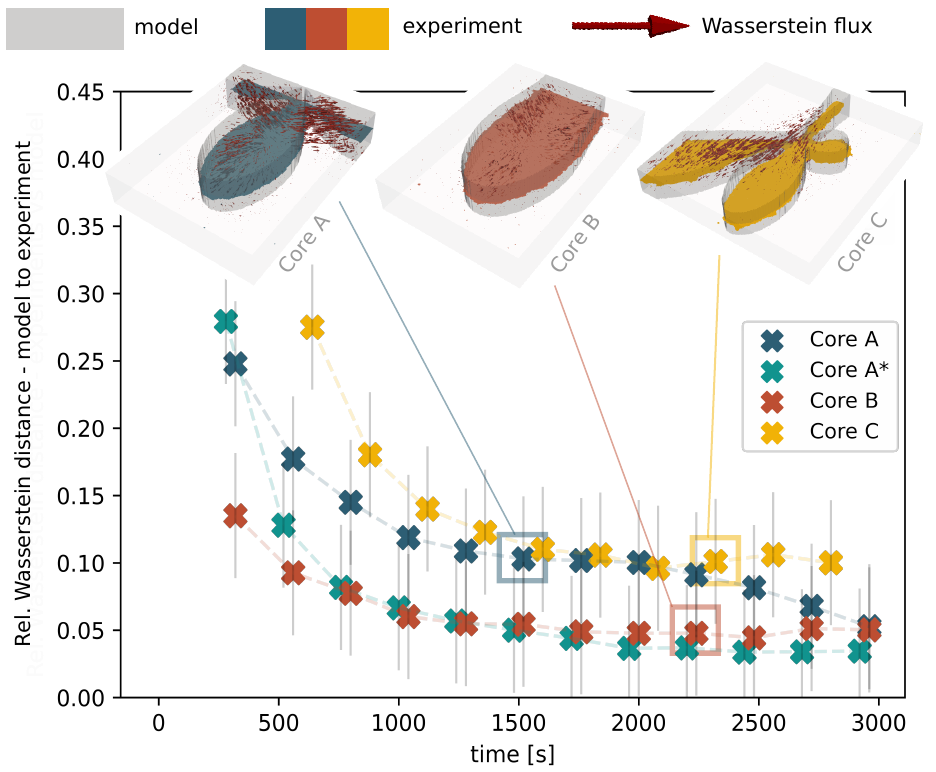} 
    \caption{Wasserstein distances over time comparing model against respective experiments; mean of measuring uncertainty indicated by error bars; Wasserstein fluxes illustrated for single time steps.}
    \label{figure:wasserstein-distances}
\end{figure}

A qualitative visual comparison see Fig.~\ref{fig:qualitative-comparison-late-time} supports the assessment of a close agreement between the modeling and experimental results. The discrepancies can be summarized as follows: For all cores, slight dispersive over- and undershoots can be observed at single locations along the front (highlighted) which are not reproduced by the simulations. However, for all cases with a single fracture (Core A/A$^\star$ and Core B, being equivalent to Core A apart from the boundary condition), the general structure of the concentration distribution is comparable between simulation and experiment, with the largest discrepancies for Core A/A$^*$ near the inlet boundaries, indicating that the realization of the boundary conditions may be of greater significance than the modeling of the fracture tip. This aligns with quantitative of Cores A/A$^*$ and B, cf. Fig.~\ref{figure:wasserstein-distances}, where Core B, with the simplest boundary condition, has the closest fidelity between experiment and simulation.

\textit{Implicit assessment of impact of fracture intersections} Fracture intersections have a wealth of structural parameters which prohibit a full direct comparison in the experimental setting. By constructing a single core with a non-orthogonal crossing, we therefore must rely on more implicit arguments. We first return to the qualitative impression gained from Fig.~\ref{fig:qualitative-comparison-late-time}. The impression is that all the main features of the tracer distribution associated with Core C are reproduced by the simulation, including the asymmetric flow out of the fractures and the non-trivial distribution of tracer near the intersection and the fracture tips. The time chosen in Fig.~\ref{fig:qualitative-comparison-late-time} is representative; other times are provided in supplemental material~\cite{Both2023supp}. To support this qualitative comparison, Wasserstein distances are again computed between the regularized laboratory data and corresponding simulation data, cf.\ Fig.~\ref{figure:wasserstein-distances}. The relative distances again substantiate the qualitative comparison above, with the discrepancy between simulation and model for Core C being within a factor 2 of the other cores, despite the increase in geometric complexity. 
Inspection of associated Wasserstein fluxes (insets) allows for interpretation of the deviations, in particular emphasizing the unbalanced fluid distribution among the fractures in Core C. Based on the available data, it is not possible to determine if these deviations are due to experimental imperfections or the modeling. 

% \section{Conclusion}
In conclusion, we return to our primary objective, that is to by means of careful experiments assess the validity of mixed-dimensional modeling of fractured porous media. Overall, across all four experimental setups we see a relative error on the order of 10\% between simulation and model, which in the context of the complexity associated with transport in fractured porous media~\cite{berkowitz2002characterizing}, we consider very satisfactory. As such, our overall assessment is that the data and simulations presented herein provide a strong experimental justification for the applicability of mixed-dimensional modeling concepts in fractured porous media.

% Bibtex references
% \bibliography{references} 

%%%%%%%%%% Merge with supplemental materials %%%%%%%%%%
\newpage\newpage
\pagebreak
\widetext
\begin{center} 
\textbf{\large Supplemental Material: High-fidelity experimental model verification for flow in fractured porous media}
% \maketitle 
\end{center}
\setcounter{equation}{0}
\setcounter{figure}{0}
\setcounter{table}{0}
\setcounter{page}{1}
\setcounter{section}{0}
\nocite{*}
\makeatletter
\renewcommand{\thefigure}{S\arabic{figure}}

This supplemental material for~[1] is three-parted and contains I. material parameters of the considered tracer experiments, as well as II. additional figures substantiating the observations and analysis of the main material, and finally III. a discussion on the use of relative Wasserstein distances accompanied by a statistical analysis of the regularization and experimental errors asserting the high quality of the experimental data of the study. For II, additional snapshots of side-by-side comparisons are provided for visual comparison of the shape of plumes for the different experiments, complemented by illustrations of the Wasserstein fluxes indicating the quality of their match. It is important to emphasize that the conclusions from the main material equally apply to the supplemental material, which thereby merely provides a fuller account of the discussion in the main material~[1].

\section{Material parameters}

The three considered core samples (Core A/A$^\star$, B, C) are treated homogeneous and isotropic. Based on independent displacement experiments using intact core material, effective hydraulic matrix properties (porosity, permeability, dispersion) are determined cf.\ Table~\ref{tab:parameters}.

\begin{table}[hpt!] 
    \centering
    \begin{tabular}{l c c c}
    \hline
     Prop. [unit] & Core A/A$^\star$ & Core B & Core C\\
     $\phi$  &  0.232 & 0.221 &  0.211 \\
     $K_{||}$ [D] & 1.9 $\pm$ 0.1 & 1.8 $\pm$ 0.07 & 1.5 $\pm$ 0.1  \\
     $\alpha_\mathrm{L}$ [m] & 7e-4 & 7e-4  & 7e-4 \\
     $\alpha_\mathrm{T}$ [m] & 0.2 $\alpha_\mathrm{L}$ & 0.2 $\alpha_\mathrm{L}$ & 0.2 $\alpha_\mathrm{L}$ \\
         \hline
    \end{tabular}
    \caption{Material properties at ambient dimension.
    \label{tab:parameters}}
\end{table}

\section{Additional snapshots substantiating the model validation}

\begin{figure}[!h]
    \centering
    \includegraphics[width=0.95\textwidth]{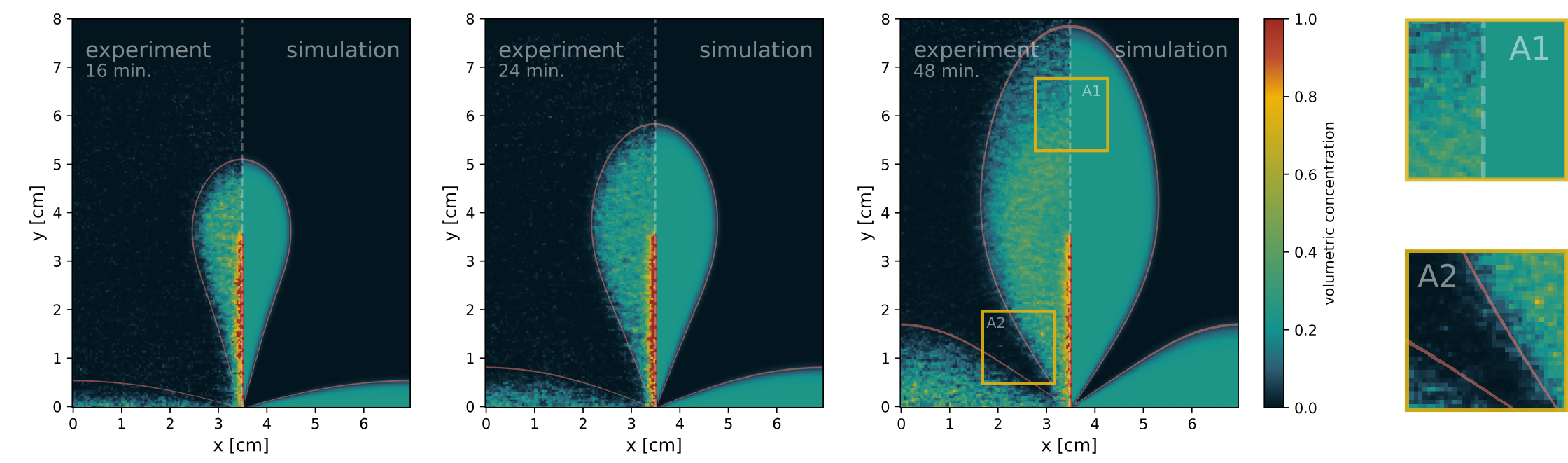}
     \caption{Core A: Evolution of the (essentially symmetric) tracer concentration plumes for the experimental results in side-by-side comparison to the corresponding numerical simulation; additional 5\% contour line to aid visual comparison. Close-up visualisation of two characteristic regions with dispersive under- and overshoots. \label{fig:qualitative-comparison-a}}
\end{figure}

\begin{figure}[!h]
    \centering 
    \includegraphics[width=0.95\textwidth]{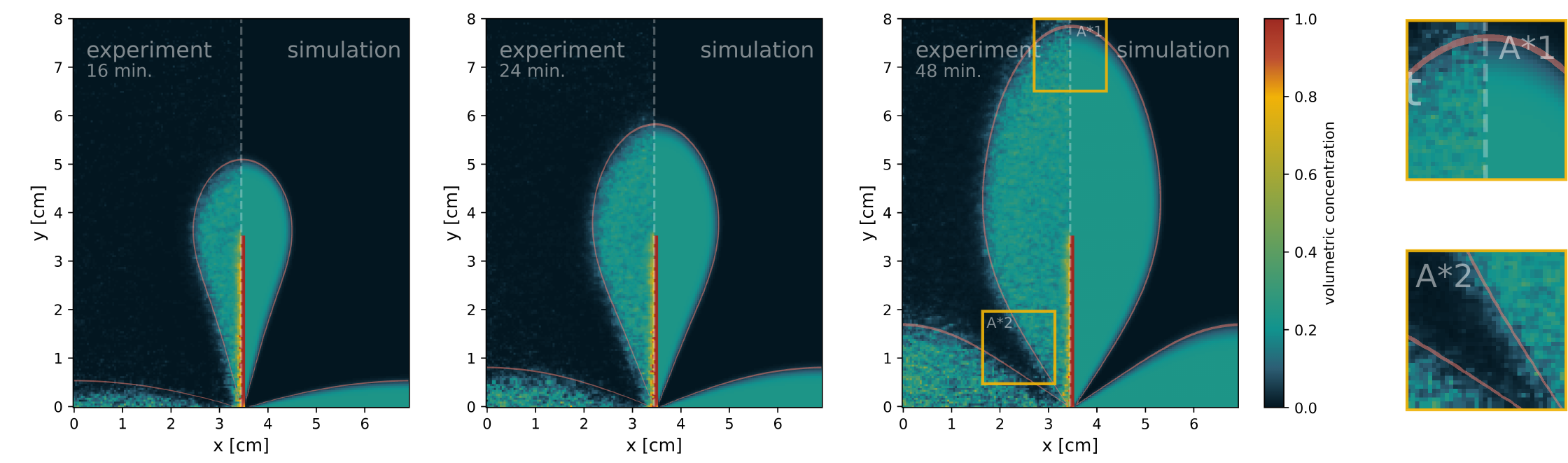}
     \caption{Core A$^\star$: Evolution of the (essentially symmetric) tracer concentration plumes for the experimental results in side-by-side comparison to the corresponding numerical simulation; additional 5\% contour line to aid visual comparison. Close-up visualisation of two characteristic regions with dispersive under- and overshoots. \label{fig:qualitative-comparison-a-star}}
\end{figure}

\begin{figure*}[!h] 
\centering
 \includegraphics[width=0.95\textwidth]{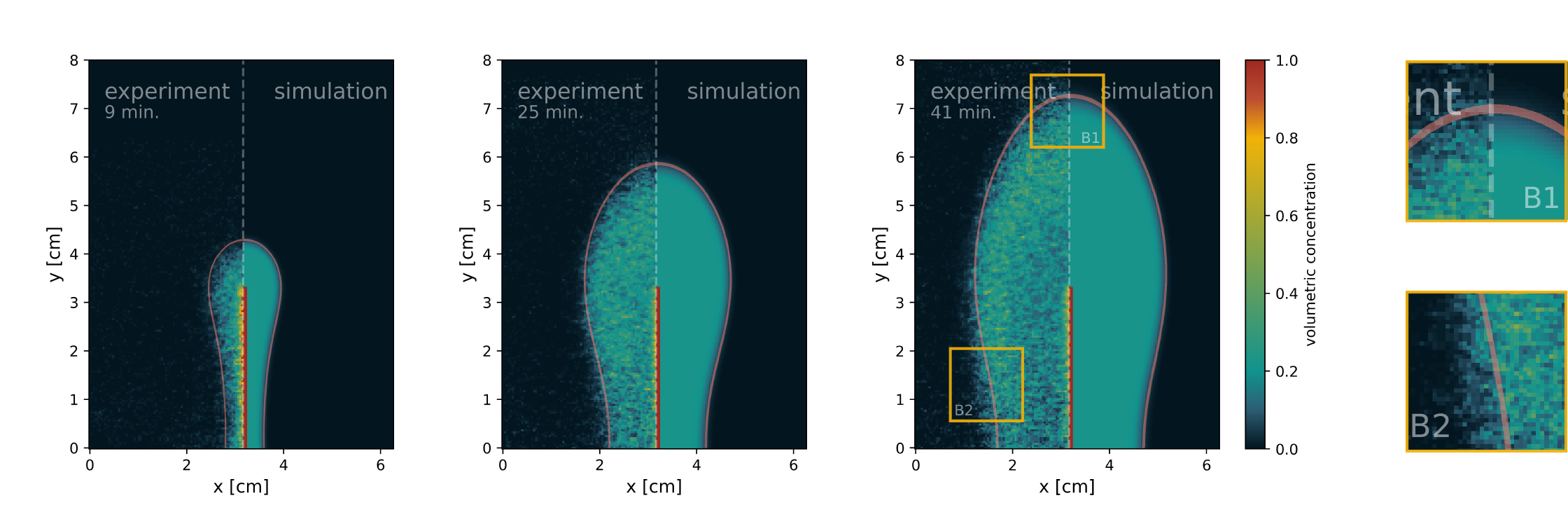}
 \caption{Core B: Evolution of the (essentially symmetric) tracer concentration plumes for the experimental results in side-by-side comparison to the corresponding numerical simulation; additional 5\% contour line to aid visual comparison. Close-up visualisation of two characteristic regions with dispersive under- and overshoots. \label{fig:qualitative-comparison-b}}
\end{figure*}

\begin{figure}[!h] 
    \centering
    \includegraphics[width=0.8\textwidth]{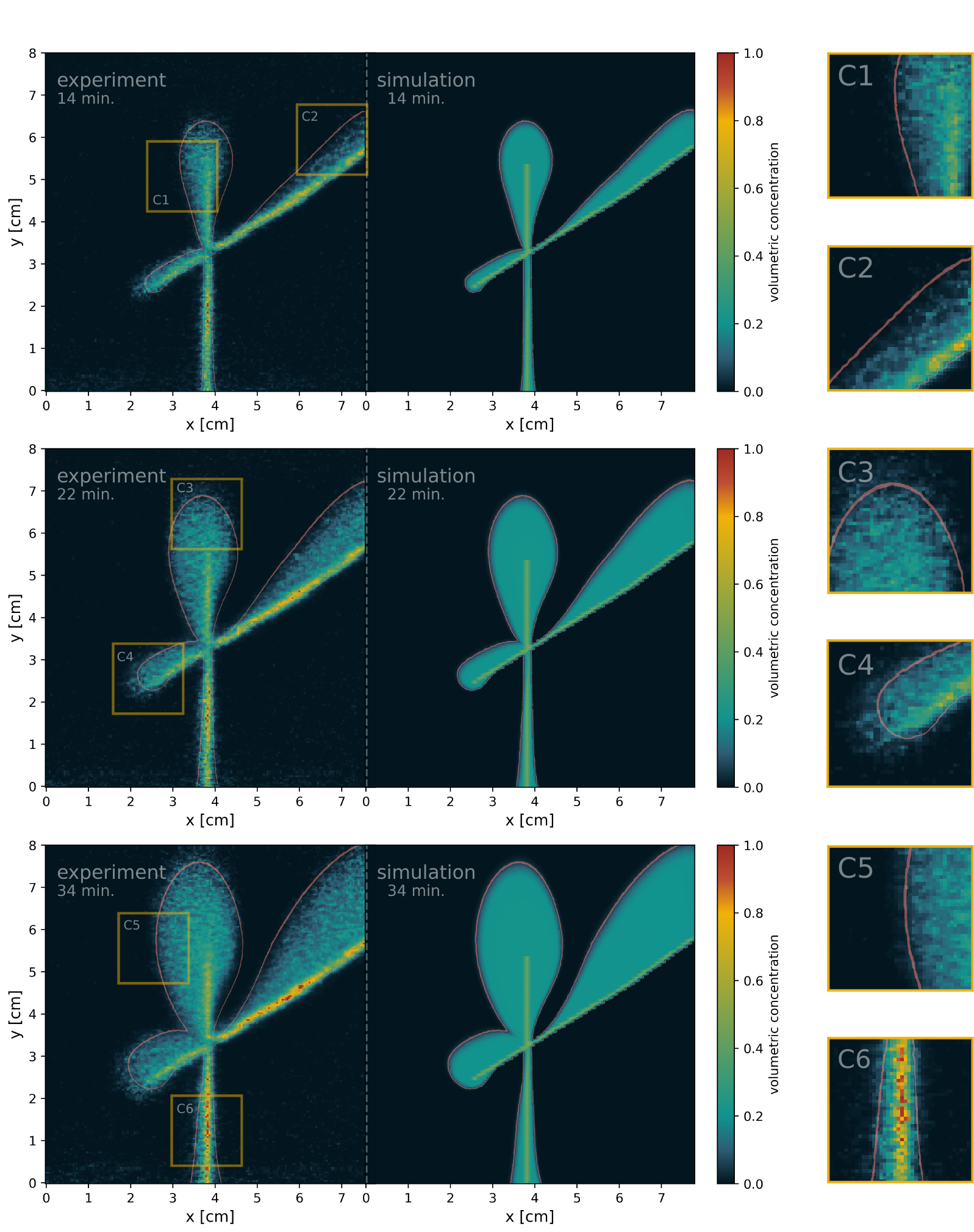} 
     \caption{Core C: Evolution of the (essentially symmetric) tracer concentration plumes for the experimental results in side-by-side comparison to the corresponding numerical simulation; additional 5\% contour line to aid visual comparison. Close-up visualisation of two characteristic regions with dispersive under- and overshoots. \label{fig:qualitative-comparison-c}}
\end{figure}

Side-by-side comparisons are a central tool in~[1] to assess the quality of the match of model and experiments. As in the aforementioned qualitative analysis, three snapshots of each fluid displacement experiment are put side-by-side to the simulation results. Marked regions with auxiliary zoom-ins highlight some of the visually strongest discrepancies. The development of the fronts for Cores A, A$^\star$, B, and C is displayed in Fig.~\ref{fig:qualitative-comparison-a}, Fig.~\ref{fig:qualitative-comparison-a-star}, Fig.~\ref{fig:qualitative-comparison-b}, and Fig.~\ref{fig:qualitative-comparison-c}, respectively. The additional snapshots are in accordance to the presented overview in~[1], supporting the there presented conclusions.

\newpage\newpage

\begin{figure*}[h!] 
\centering
 \includegraphics[width=\textwidth]{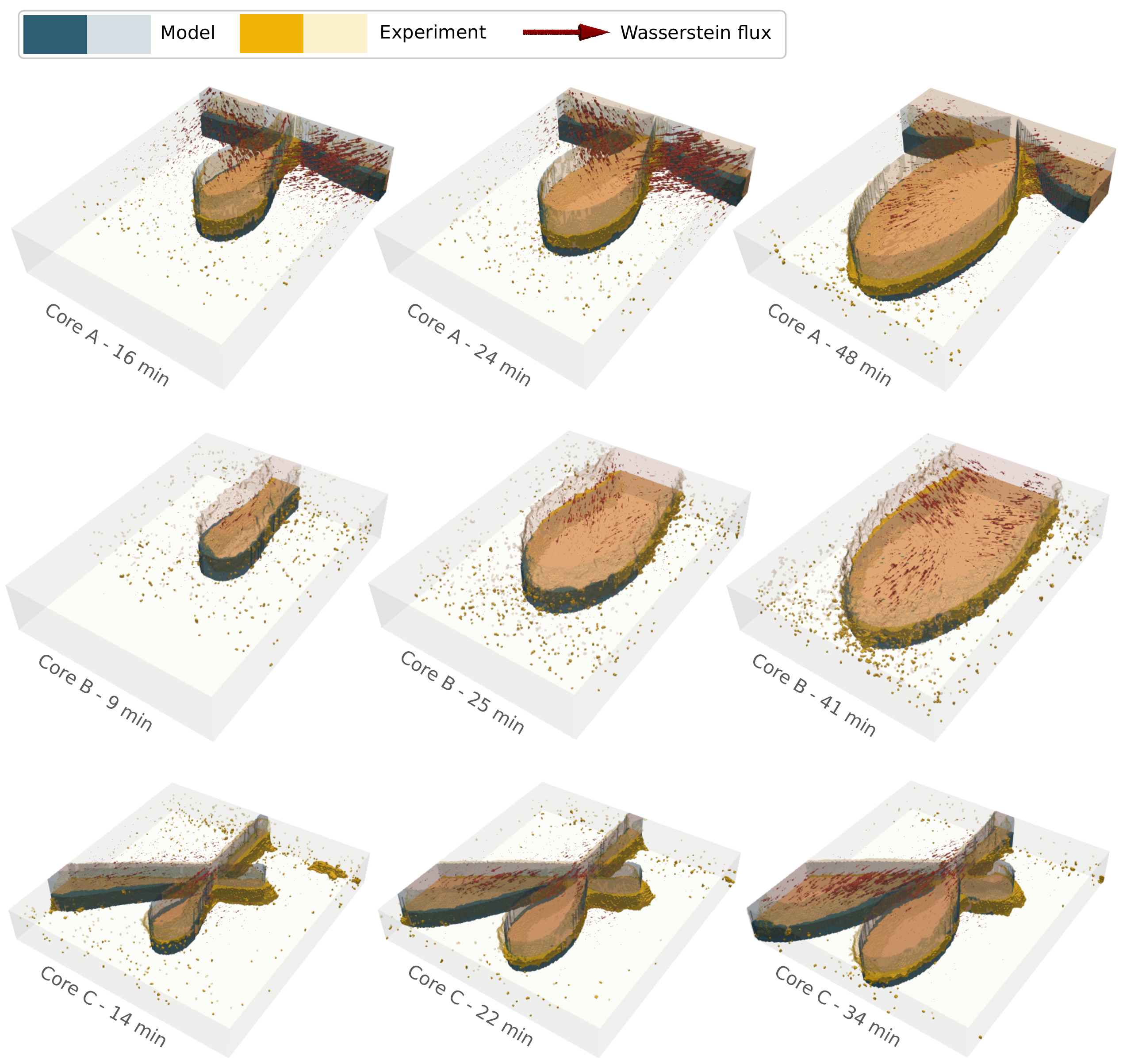}
 \caption{Cores A, B, C: Three-dimensional qualitative comparison of tracer plumes corresponding to Fig.~\ref{fig:qualitative-comparison-a},~\ref{fig:qualitative-comparison-b},~\ref{fig:qualitative-comparison-c}. Plumes at upper bottom displayed transparent. Additionally, scaled Wasserstein fluxes displayed to convert one distribution into the other. \label{fig:wasserstein}}
\end{figure*}

The above visual material is supported by illustration of Wasserstein fluxes, i.e., a visual account of the Wasserstein distance, cf.\ Sec.~\ref{sec:wasserstein}. In Fig.~\ref{fig:wasserstein}, a closer view at several Wasserstein fluxes is provided corresponding to the same time steps as considered above. In addition, both the experimental and numerical three-dimensional data is plotted on top of each other, demonstrating a good fit in terms of their shapes.

\section{Wasserstein metric entering the quantitative analysis and error quantities\label{sec:wasserstein}}

\subsection{Relative Wasserstein distance}
\begin{figure}[!h]
 \centering
 \begin{overpic}[width=0.7\textwidth]{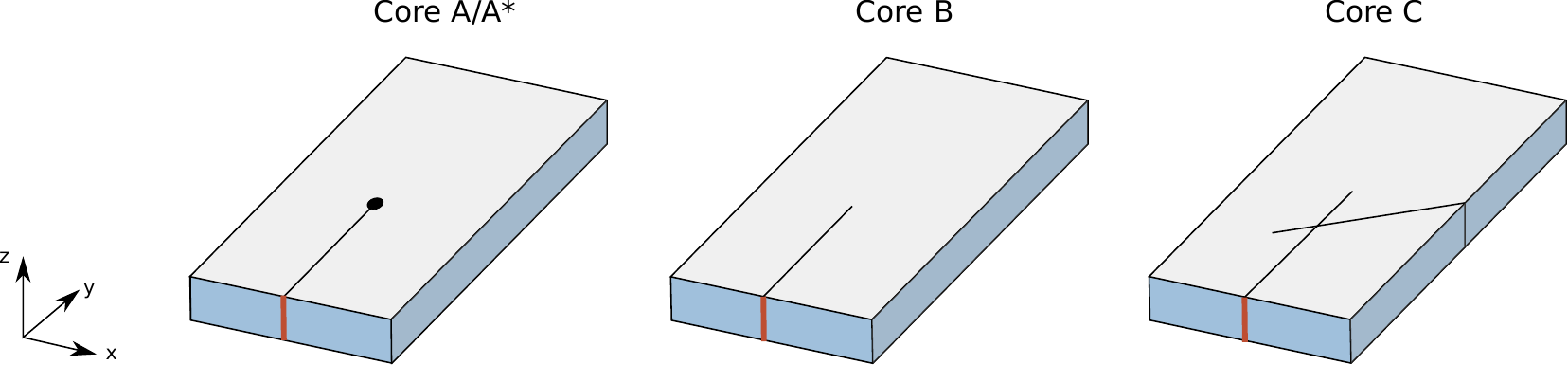}  
  \put(14.5,-1){\textcolor{orange2}{$\Gamma$}}
  \put(45,-1){\textcolor{orange2}{$\Gamma$}} 
  \put(75.5,-1){\textcolor{orange2}{$\Gamma$}}
 \end{overpic}
 \caption{\label{fig:inflow-boundary} Support $\Gamma$ for Dirac variants of Cores A/A$^\star$, B, and C, indicated in red color; cf. Fig.~2 in~[1] for full geometrical specifications.} 
\end{figure}

The quantitative analysis presented in the main material~[1] is based on Wasserstein distance measuring the difference between upscaled laboratory data and simulation data. We recall a standard definition of the 1-Wasserstein distance in terms of the solution of a variational problem also called Beckmann problem~[3]. For two compatible concentration profile $c_\mathrm{A}$ and $c_\mathrm{B}$, defined over the domain $\Omega$ with $\int_\Omega c_\mathrm{A} dx = \int_\Omega c_\mathrm{B} dx$, we define the 1-Wasserstein distance to be
\begin{align*}
 W^1(c_\mathrm{A}, c_\mathrm{B}) = \mathrm{inf} \left\{ \int_\Omega |q| dx \ : \ \nabla \cdot q = c_\mathrm{A} - c_\mathrm{B}\ \text{in }\Omega \right\}.
\end{align*}
The associated flux $q$ is also termed \textit{Wasserstein flux} in~[1]. To compute $W^1(c_\mathrm{A}, c_\mathrm{B})$ for voxel distributions $c_\mathrm{A}$ and $c_\mathrm{B}$, we use a numerical approximation~[2]. To put the results of the quantitative analysis in context, relative distances have been used. Using the average transport distance as reference value we consider the relative distance between two concentration profiles $c_\mathrm{A}$ (typically experimental data) and $c_\mathrm{B}$ (typically simulation data)
\begin{align*} 
 \text{relative Wasserstein distance} = \frac{W^1(c_\mathrm{A}, c_\mathrm{B})}{W^1(c_\mathrm{A}, \delta_{c_\mathrm{A},\Gamma})},
\end{align*}
where $\delta_{c_\mathrm{A},\Gamma})$ denotes a Dirac-type concentration variant of $c_\mathrm{A}$ with same total mass but concentrated to some part of the domain $\Gamma$; i.e., $\delta_{c_\mathrm{A},\Gamma})$ has support in $\Gamma$, is constant on $\Gamma$ and satisfies $\int_\Omega \delta_{c_\mathrm{A},\Gamma}) dx = \int_\Omega c_\mathrm{A} dx$. Here, we choose $\Gamma$ to be the intersection of the inlet and the connected fracture, cf. Fig.~\ref{fig:inflow-boundary}. By design the relative Wasserstein distance is unit-free. Values of order 1 are considered to describe order 1 deviation from the reference data (here $c_\mathrm{A}$).

\subsection{Regularization error}
The upscaled data is obtained through regularization of the raw, to large extent sparse, and noisy PET signal. As the used regularization (total variation denoising) enables inpainting, the Wasserstein distance between the raw and regularized signals, rescaled to same mass, identifies the cost required to distribute the sparse signal. This distribution is of local nature and thus identifies a small reference distance. The \textit{regularization error} given by the Wasserstein distance between raw and regularized data for the four experiments (Cores A, A$^\star$, B, and C) are displayed in Fig.~\ref{fig:regularization-error}; we employ relative distances with the reference value defined through the regularized data, compatible with the quantitative analysis in~[1]. We conclude that the relative regularization error is of the order 0.05 $\pm$ 0.03, which allows to conclude that the regularization error is of low order.

\subsection{Experimental variability measure}
\begin{figure}[h!]
 \centering
  \subfloat[Regularization error\label{fig:regularization-error}]{\includegraphics[width=0.5\textwidth]{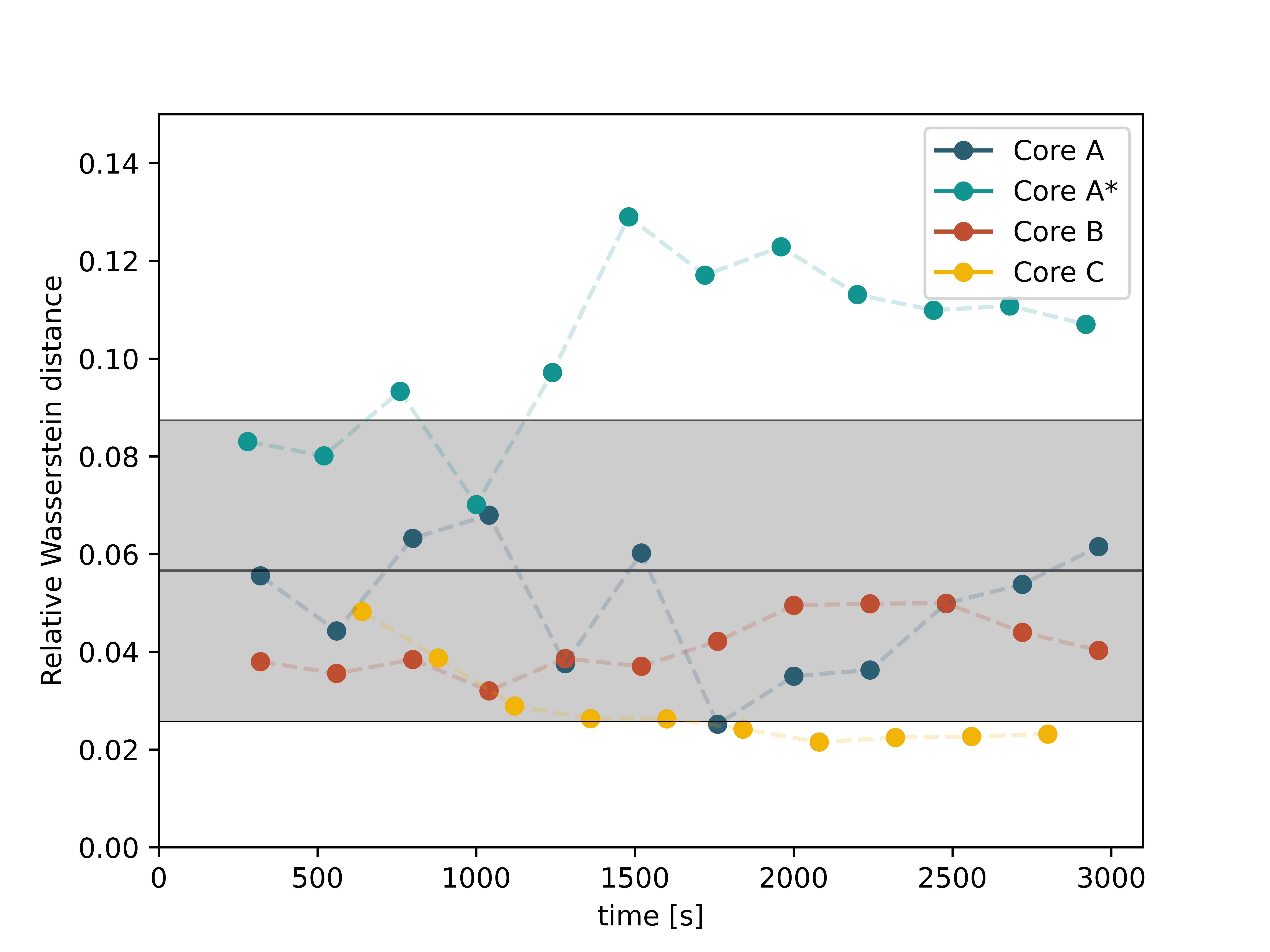}}
  \subfloat[Symmetry error\label{fig:symmetry-error}]{\includegraphics[width=0.5\textwidth]{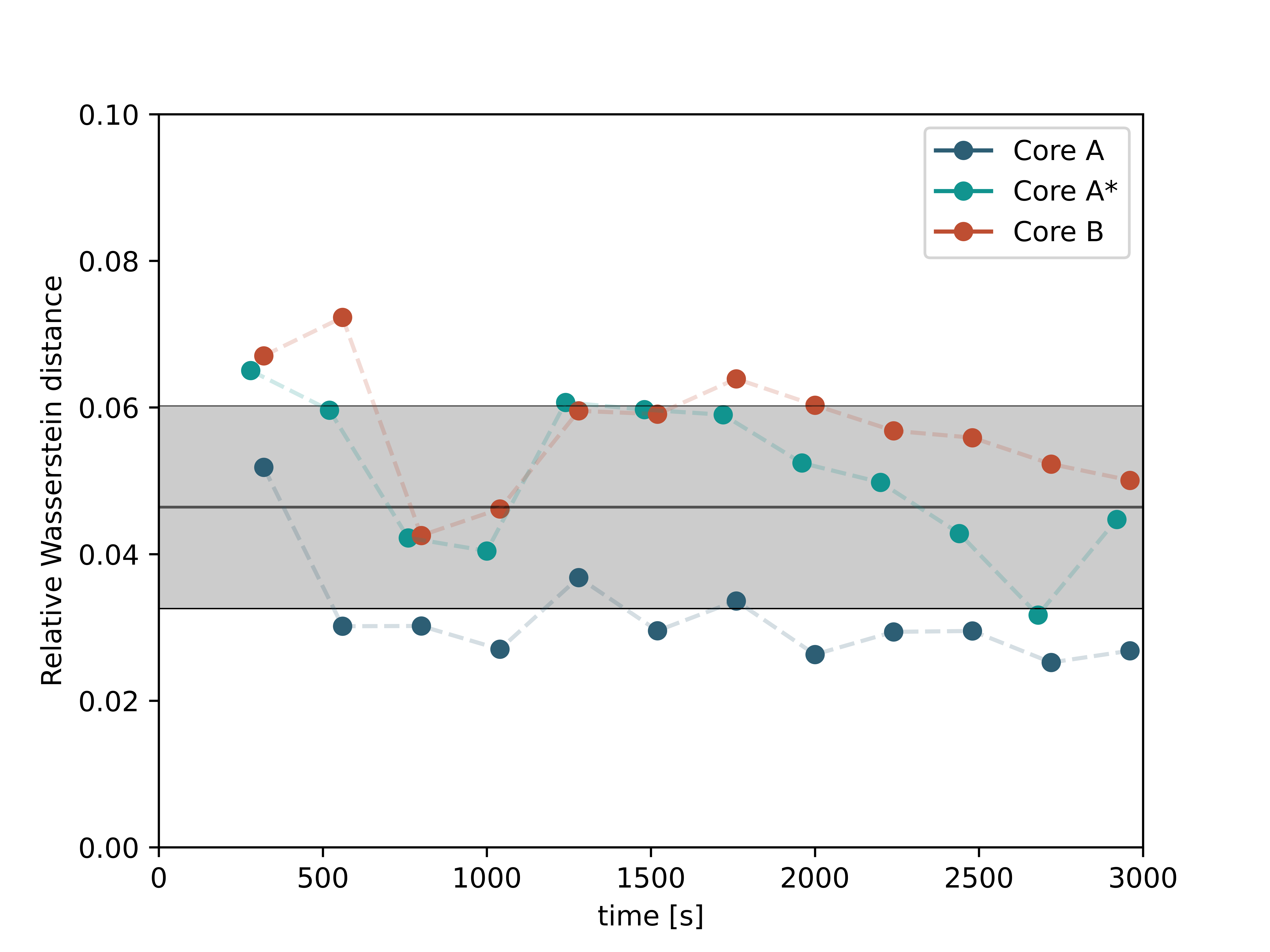} }
 \caption{Relative regularization and symmetry error for all experiments, over all time steps; mean and standard deviation illustrated by gray box.}
\end{figure}

The experiments for Cores A, A$^\star$ and B are designed to be plane symmetric. Yet, naturally, experiment imperfections from merely nearly isotropic conditions and influences (e.g. operating conditions) affect the fluid displacement. After all, the resulting fluid displacement is not idealistically symmetric. We exploit that fact to quantify the degree of variability. For this we quantify asymmetry by measuring the Wasserstein distance of the regularized experimental data sets and its mirrored image, across the presumed North-South symmetry plane (along the fracture). The resulting \textit{symmetry error} for the experiments with symmetric character is displayed in Fig.~\ref{fig:symmetry-error}; as for the regularization error and compatible with~[1], we employ relative distances with the reference value defined through one of the two regularized concentrations.  We conclude that the relative symmetry error is of the order 0.046 $\pm$ 0.013, which asserts high quality of the data. Due to the identical operating conditions, it is assumed that the symmetry error is also representative for experimental data for Core C.

% \newpage 

\begin{tabular}{ll}
$[1]$ &\textit{J.\ W.\ Both, B.\ Brattekås, E.\ Keilegavlen, M.\ A.\ Fernø, and J.M.\ Nordbotten:} High-fidelity experimental\\
& model verification for flow in fractured porous media, \textit{submitted}. \\
$[2]$ &\textit{J.\ W.\ Both, E.\ Facca, and J.\ M.\ Nordbotten:} Iterative
finite volume approximation of the 1-Wasserstein\\
& distance, \textit{in preparation}. \\
$[3]$& \textit{F.\ Santambrogio:} Optimal transport for applied mathematicians, \textit{Birk{\"a}user, NY 55.58-63: 94,} 2015. 
\end{tabular}

\end{document}